\documentclass[epj]{svjour}

\usepackage{graphicx}
\usepackage{fancyhdr}
\def\makeheadbox{{%  remove EPJC header
 }}
\def\mytitle{My title} 
\def\myauthors{My name}  
\def\mytype{My type of session}
\def\mysession{My session}
\def\mytitle{Flavour Violating Interactions of Supersymmetric Particles} %Put your title here!
\def\myauthors{Andrew Box}    %Put your name here!
\def\mytype{Contributed Talk}    
\def\mysession{Colliders - SUSY Phenomenology}
\newcommand{\sn}{s}
\newcommand{\cs}{c}

\newcommand{\gtpur}{(\tilde{g}'_{u_R})}
\newcommand{\ftuq}{(\tilde{f}^Q_u)}
\newcommand{\ftur}{(\tilde{f}^{u_R}_u)}
\newcommand{\gtphu}{(\tilde{g}'_{h_u})}
\newcommand{\gtpq}{(\tilde{g}'_Q)}

\newcommand{\h}{\theta_h}
\newcommand{\Hh}{\theta_H}
\newcommand{\sh}{\theta_{\tilde{h}}}
\newcommand{\sqk}{\theta_{{\tilde{Q}_k}}}
\newcommand{\sbi}{\theta_{\tilde{B}}}
\newcommand{\suk}{\theta_{{\tilde{u}_k}}}

\newcommand{\by}{\mathbf{Y}}
\newcommand{\bu}{\mathbf{U}}
\newcommand{\bv}{\mathbf{V}}
\newcommand{\bw}{\mathbf{W}}
\newcommand{\bx}{\mathbf{X}}

\begin{document}
\title{Flavour Violating Interactions of Supersymmetric Particles}
\author{Andrew D. Box
}                     % Do not remove
\institute{Department of Physics and Astronomy, University of Hawaii, Honolulu, HI, 96822, USA}
\date{}
\abstract{
We derive the renormalization group equations for dimensionless couplings and soft supersymmetry breaking (SSB) parameters given arbitrary high scale inputs, incorporating 1-loop thresholds. A program (to be incorporated into ISAJET) has been developed, which includes flavour violating couplings of superparticles based on an arbitrary framework for the high scale flavour structure of SSB parameters. The weak scale flavour structure derived in this way can be applied to the study of SUSY flavour changing decays. As an illustration, we recalculate the branching ratio of the flavour-violating decay of the top squark.
 % end of PACS codes
} %end of abstract
\maketitle
%DO NOT REMOVE THIS LINE
%

\section{Introduction}
\label{intro}

In most supersymmetric models, the soft SUSY breaking parameters are determined by high scale physics. Using an appropriate high scale ansatz, weak scale SUSY couplings relevant for phenomenology can be calculated using the renormalization group equations (RGEs). This procedure also yields predictions for flav-our violation at the weak scale, assuming of course we have a theory (or at least an ansatz) of flavour at a high energy scale.

To accurately obtain the RG prediction at the two loop level, effects of thresholds at one loop must be incorporated. These include the decoupling of heavy particles and the concomitant splitting between couplings that are equal in the SUSY limit (the gauge and gaugino couplings, for instance). These new couplings will be identified with a tilde and labelled with the scalar which appears in the operator, e.g. the $\tilde{B}u\tilde{u}_{R}$ coupling is denoted by $\gtpur$.

The fact that gaugino couplings can become distinct from their corresponding SM gauge counterparts means that they can even develop flavour off-diagonal terms. These couplings therefore play an important role in flavour physics. It is also important to remember that threshold effects must be included in the running of the soft masses and trilinear SSB parameters.

The RGEs are constructed to describe a collection of effective theories (with varying particle content) valid at different scales. We approximate the changing particle content via step functions which remove the influence of the heavy particles at the appropriate scales \cite{cas2,dedes}. Each threshold is denoted by a $\theta$ and labelled with the relevant particle. Above all thresholds, in the unbroken regime, all $\theta$-functions are unity, and the running is as given by the MSSM \cite{martv}. Below all sparticle and additional Higgs thresholds, these step functions all vanish and the running is purely SM \cite{cas1}.

The importance of considering these effects can be shown by re-examining previous estimates for the flavour changing decay $\tilde{t}_{1}\rightarrow c\tilde{Z}_{1}$, which requires a solution of the RGEs in order to estimate $\tilde{t}-\tilde{c}$ mixing. This has previously been estimated \cite{HK} by integrating the RGEs in a single step. We will show that the results are significantly different when the RGEs are solved numerically, and the various effects discussed above are taken into account.

\section{Deriving and solving the threshold RGEs} \label{sec:der}

The general procedure for inserting thresholds begins with RGEs for a general theory \cite{machv} which do not depend on the particle structure of the theory. The RGEs in \cite{machv} are written for dimensionless parameters. In \cite{luo} they are rewritten for complex fermions and expanded to include RGEs for dimensionful parameters which are necessary for supersymmetric theories. We write the interaction of 2-component Majorana fermions ($\psi_{p}$) with real scalars ($\phi_{a}$) via Yukawa couplings ($Y^{a}_{pq}$) as
\begin{equation} \label{eq:2complag}
\mathcal{L}_{int}=-\left(\frac{1}{2}Y^{a}_{pq}\psi^{T}_{p}\zeta\psi_{q}\phi_{a}+\mathrm{h.c.}\right),
\end{equation}
using $\zeta$ as the metric which makes the fermion bilinear Lorentz invariant. With this interaction, the correct form of the general Yukawa RGE is
\begin{eqnarray}
\nonumber\lefteqn{\left(4\pi\right)^2\left.\frac{d\by^{a}}{dt}\right|_{1-loop}=}\qquad\qquad \\
\nonumber&&\frac{1}{2}\left[\by^T_2(F)\by^a+\by^a\by_2(F)\right]+2\by^b\by^{\dagger a}\by^b\\
\nonumber&&+\by^b\mathrm{Tr}\left\{\frac{1}{2}\left(\by^{\dagger b}\by^a+\by^{\dagger a}\by^b\right)\right\}\\
&&-3g^2\left\{\mathbf{C}_2(F),\by^a\right\}, \label{eq:2comprge}
\end{eqnarray}
in which $\by_2(F)=\by^{\dagger b}\by^b$ and $\mathbf{C}_2(F)=\mathbf{t}^A\mathbf{t}^A$. An example, rewritten in the more phenomenologist friendly 4-component language with complex scalars and both Majorana and Dirac fermions, is shown in Appendix~\ref{sec:4comp}. It should be stressed that though the equations look much lengthier when written in 4-component form, much work needed to derive the RGEs has already been done, so their larger size is also indicative of ease of use.

To facilitate decoupling, interactions of the fields need to be written in the (approximate) mass basis. In the Higgs sector the real neutral fields $h$, $G^{0}$, $H$ and $A$ are combined into two complex Higgs fields, $\mathsf{h}$ and $\mathcal{H}$ (using a different font to differentiate the complex fields from the real fields) such that
\begin{eqnarray}
\mathsf{h}&=\frac{h+iG^{0}}{\sqrt{2}}\\
\mathcal{H}&=\frac{-H+iA}{\sqrt{2}}.
\end{eqnarray}
$\mathsf{h}$ and $\mathcal{H}$ along with the complex charged fields $G^{+}$ and $H^{+}$, are obtained from the usual Higgs doublets (in the notation of \cite{wss}) through a field rotation:
\begin{eqnarray}
\label{eq:hrot}\left(\begin{array}{c}G^{+}\\\mathsf{h}\end{array}\right)&=\sn\left(\begin{array}{c}h^{+}_{u}\\[5pt]h^{0}_{u}\end{array}\right)+\cs\left(\begin{array}{c}h^{-*}_{d}\\[5pt]h^{0*}_{d}\end{array}\right)\\[5pt] 
\label{eq:Hrot}\left(\begin{array}{c}H^{+}\\\mathcal{H}\end{array}\right)&=\cs\left(\begin{array}{c}h^{+}_{u}\\[5pt]h^{0}_{u}\end{array}\right)-\sn\left(\begin{array}{c}h^{-*}_{d}\\[5pt]h^{0*}_{d}\end{array}\right),
\end{eqnarray}
where $\sn=\sin{\beta}$ and $\cs=\cos{\beta}$. We consider the effects of Higgs boson thresholds in the approximation $m_{A}>>M_{Z}$, in which case $m_{H}\sim m_{A}\sim m_{H^{+}}$, with $m_{h}\sim M_{Z}$\footnote{If instead $m_{A}\sim M_{Z}$, all Higgs bosons have a roughly common mass, and threshold effects in this sector are unimportant.}. The doublet in (\ref{eq:Hrot}) then decouples at a scale $Q\sim m_{H}$ while that in (\ref{eq:hrot}) remains in the low energy theory.

For the neutralino and chargino thresholds, we ignore gaugino-Higgsino mixing (an excellent approximation if either of the conditions $\left|\mu\right|>>M_W$ or $\left|M_{1,2}\right|>>M_W$ are satisfied\footnote{If neither of these conditions are met, the gauginos and Higgsinos will all have similar mass so they will all decouple at about the same scale. It will therefore still be approximately valid to use a single scale to decouple the Higgsinos.}) so that the Higgsino mass eigenstates
\begin{equation}
\frac{\psi_{h_{d}}\pm\psi_{h_{u}}}{\sqrt{2}}
\end{equation}
are decoupled at the scale $\left|\mu\right|$, while the bino and wino states are decoupled at $Q=M_{1}$ and $Q=M_{2}$ respectively.

Below $Q\sim m_{H}$ the theory contains just the SM Higgs scalar (although it may still contain Higgsinos), and the MSSM Yukawas ($f$) are replaced by their SM equivalents ($\lambda$) using the conditions: $\lambda_{u}=\sin{\beta}f_{u}$ and $\lambda_{\{d,e\}}=\cos{\beta}f_{\{d,e\}}$. Note that if $\left|\mu\right|<m_{H}$, Higgsinos would still couple fermions to sfermions, an example of which is the term $\Psi_{h^{0}_{u}}u\tilde{u}_{R}$ with coupling $\ftur$.

The general approach to solving the RGEs is as follows:
\begin{itemize}
\item Begin with weak scale gauge couplings and quark masses. Rotate the Yukawas to the current basis using four rotation matrices chosen to make sure the KM matrix is correct.
\item Run the gauge couplings and Yukawas up to the high scale.
\item Input the high scale ansatz which can depend on the Yukawa flavour stucture,
\item Run the RGEs down in the presence of thresholds. As various thresholds are passed, remove the particles from the theory, with special attention given to the matching at $m_{H}$ as mentioned above.
\end{itemize}
Since the location of the thresholds is not known at the beginning of the process, it is necessary to begin with an estimate of their location and iterate this procedure until the solution converges to the required accuracy.

In this manner, it is possible to obtain the whole set of couplings at a scale appropriate to the problem -- for example $Q=m_{\tilde{t}_{1}}$ in the case of stop decay. The inputs are: weak scale gauge couplings and squark masses; and high scale SUSY parameters governed by a particular model.

\section{Additional Couplings when SUSY is broken} \label{sec:tilde}

When SUSY is broken by the decoupling of one or more of the SUSY particles, there are many couplings which, although equal in the supersymmetric limit, have different RGEs. This causes the couplings themselves to become different. As a simple example we can compare the quark-quark-gauge coupling to the squark-quark-gaugino coupling. These couplings are equal when SUSY is exact, but can be conceptually and numerically different below the scale where some sparticles have decoupled from the theory.

To one loop, the running of the gauge coupling does not depend on any other couplings and the only effect of thresholds is to change the numerical factor in the RGE. On the other hand, the running of the gaugino coupling contains many extra terms, some depending on various gaugino couplings and others depending on Yukawa matrices.

The extra Yukawa terms in the right-handed up-type gaugino coupling running are:
\begin{eqnarray}
\nonumber&1.\quad&\left[\sn^2\h+\cs^2\Hh\right](f_u)^T_{ik}(f_u)^*_{kl}\gtpur_{lj}\\
\nonumber&2.\quad&-3\left[\sn^2\h+\cs^2\Hh\right]\sbi\sh(f_u)^T_{ik}\ftur^*_{kj}\gtphu\\
\nonumber&3.\quad&\sh\sqk\ftuq^T_{ik}\ftuq^*_{kl}\gtpur_{lj}\\
\nonumber&4.\quad&-\sh\sqk\ftuq^T_{ik}\gtpq_{kl}\ftur^*_{lj}\\
&5.\quad&2\sh\suk\gtpur_{ik}\ftur^T_{kl}\ftur^*_{lj}
\end{eqnarray}
In the exact SUSY limit, the extra Yukawa terms cancel and the gaugino couplings are equal to the gauge coupling so that the RGEs are identical. When SUSY is broken this is no longer the case. If the heavy Higgs particles decouple first, for example, $\theta_{H}=0$ and several terms disappear from the running. This means that the cancellation of the Yukawa terms will no longer be exact, leading to a non-zero overall Yukawa factor in the RGE.  The coupling can develop non-zero off-diagonal flavour changing terms which are not present in the gauge coupling. Thus, if the squarks and gauginos are lighter than $m_{H}$ the RGE for this coupling will have a Yukawa component at a scale where it is still important for phenomenology.

\begin{figure}
\includegraphics[width=0.348\textwidth,height=0.45\textwidth,angle=270]{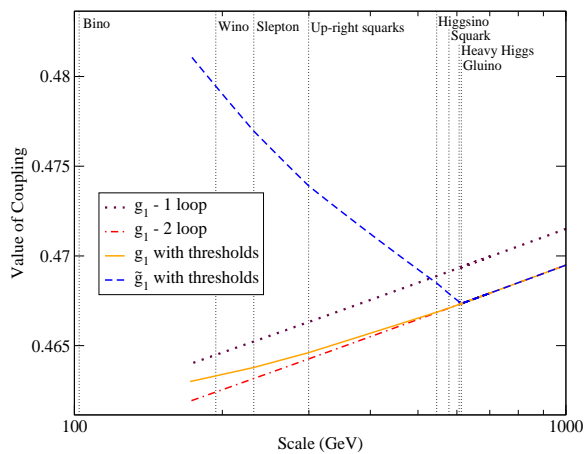}
\caption{The evolution of the U(1) gauge coupling to 1-loop order (dotted curve) and 2-loop order (dot-dashed curve). Also the U(1) gauge coupling with thresholds (solid). The upper line (dashed) is the U(1) gaugino coupling to a right-handed up-type squark ($\tilde{g}_{1}$). Vertical dotted lines indicate the presence of thresholds.}
\label{fig:ggtild}
\end{figure}

In the U(1) gaugino running, the decoupling of other gaugino terms can also result in a different value for the diagonal elements, as shown in Figure~\ref{fig:ggtild}. Here we show the 1-loop and 2-loop  running of $g_{1}$ in the dotted curve and dot-dashed curve respectively. The solid line shows the running of the $g_{1}$ coupling when thresholds are switched on. It is clear that at $Q=m_{t}$ the effect of thresholds on the coupling is comparable to the difference between the 1- and 2-loop lines. The dashed line running upwards from the gluino threshold is the $\tilde{g}'_{u_{R}}$ coupling which is important until the up-right squarks decouple, shown by the central dotted line. We can see that the gaugino coupling at this point is very different from its SM counterpart.

There are many terms in the full threshold RGEs which can provide an unexpected dependence like the Yukawa terms in the gaugino coupling RGE. Another such example is in the running of the CP-conserving gaugino mass, $M_{2}$ \cite{wss}, for which
\begin{equation}
\left(4\pi\right)^{2}\frac{dM_{2}}{dt}\ni2M_{2}\sn\cs\left(-\theta_{h}+\theta_{H}\right)\theta_{\tilde{h}}\left[\mu^{*}\tilde{g}_{h_{d}}\tilde{g}_{h_{u}}+\mathrm{c.c.}\right],
\end{equation}
It is clear that this term does not contribute in the SUSY limit (when all $\theta=1$) but can appear in the RGE if $m_{H}>\left|\mu\right|$.

We have obtained the full system of threshold RGEs for the gauge couplings and Yukawas, in addition to threshold RGEs for $\mu$, gaugino masses, soft masses and trilinear SSB parameters \cite{abfut} and plan to incorporate our code into ISAJET \cite{ijet}.

\section{Flavour changing top squark decay} \label{sec:HK}

The two body decay of the lighter stop ($\tilde{t}_{1}$) into a charm quark and the lightest neutralino ($\tilde{Z}_{1}$) occurs at the one loop level. This decay becomes important if tree level two body decays are kinematically forbidden, although it may still compete with three body decays \cite{porod}.

Two body flavour changing stop decay was first studied by Hikasa and Kobayashi \cite{HK}, where the off-diagonal elements of the up-squark SSB matrix, specifically $\tilde{t}_{\{L,R\}}-\tilde{c}_{L}$ mixing elements, were estimated under the approximation that the RGEs could be integrated using a single step and assuming the LSP was a photino\footnote{The Hikasa-Kobayashi result was later modified to allow for arbitrary composition of the LSP.}. The partial decay width can then be readily obtained. Although their estimate was adequate for the very light stop case they were interested in, it is not valid for $m_{\tilde{t}_{1}}$ in the range of interest today.

To this end, the RG method was used to calculate the mixing numerically. This removes the need for the one-step approximation and also allows us to obtain the full weak scale couplings for use in the decay calculation. The decay width was re-derived from the Lagrangian, keeping the difference between sparticle and SM couplings, which allows for the possibility of extra flavour changing terms coming from sources other than the up-squark mass matrix.

Table~\ref{tab:comp} compares this partial width calculated using the various methods. The Hikasa-Kobayashi result clearly overestimates the width, illustrating the importance of our improvement. In a random sample of 10,000 mSUGRA points, over 300 were found with a stop mass above $100\ \mathrm{GeV}$ and with kinematically forbidden two body decays. In these cases the one-step approximation overestimated the width by between a factor of 15-25 although in a few cases the difference was as large as a factor of 35. 
\begin{table}
\caption{Comparison of the calculated width ($\times10^{-9}\ \mathrm{GeV}$) for $\tilde{t}_{1}\rightarrow c\tilde{Z}_{1}$ using various methods. In all cases SSB thresholds were not used and the GUT scale inputs were the mSUGRA parameters $m_{0}=300\ \mathrm{GeV}$, $m_{1/2}=250\ \mathrm{GeV}$, $A_{0}=-1000\ \mathrm{GeV}$, $\tan{\beta}=10$, $sign(\mu)=+1$, $m_{t}=172\ \mathrm{GeV}$ which led to $m_{Z_{1}}=100\ \mathrm{GeV}$ and $m_{\tilde{t}_{1}}\simeq183\ \mathrm{GeV}$.}
\centering
\label{tab:comp}
\begin{tabular}{lc}
\hline\noalign{\smallskip}
Method & Width \\
\noalign{\smallskip}\hline\noalign{\smallskip}
Hikasa-Kobayashi & $\sim18$ \\
1-loop (all thresholds at $m_{H}$) & 1.04 \\
2-loop (all thresholds at $m_{H}$) & 1.15 \\
2-loop (realistic thresholds) & 1.19 \\
2-loop (realistic thresholds and tilde terms) & 1.31 \\
\noalign{\smallskip}\hline
\end{tabular}
\end{table}

Looking at the rest of the results in the table, we can see that introducing thresholds to the equations produces a result which, although less than the difference between the 1-loop and 2-loop running, is of similar order. This indicates that the thresholds are important for claiming true 2-loop accuracy. The introduction of tilde terms produces a sizeable change, which shows that the difference between sparticle and SM couplings is also important.

Clearly the single step integration may give a qualitatively misleading value of the $\tilde{t}_{1}$ decay branching ratios in the event there are competing modes. The other effects are also important for obtaining a quantitative projection of stop decay patterns. 

\section{Conclusions} \label{sec:conc}

We have constructed a closed system of RG equations including threshold decoupling for sparticles and heavy Higgs particles. This includes SSB trilinear couplings and masses in addition to the gauge and Yukawa couplings. A code has been developed to solve the RGEs which will be incorporated into the Isajet event generator.

When SUSY is broken, the RGEs for sparticle couplings can become different from the RGEs for their SM counterparts, i.e. the gauge and Yukawa couplings. This not only results in different diagonal elements at the weak scale, but many sparticle couplings \textit{also develop additional flavour off-diagonal terms}.

Our consideration of the flavour changing decay of the stop finds the partial width to have been over-estimated by a factor of 15 to 25 as a result of the one-step approximation. We also find that threshold effects in both the standard couplings and the so-called tilde terms are important for more quantitative flavour violation calculations.

\subsubsection*{Acknowledgements}

This research was carried out in collaboration with Xerxes Tata and was supported in part by a grant from the US Department of Energy. The University of Hawaii Graduate Student Organisation provided partial support for accommodation during the Conference.

\appendix
\section{4-component Yukawa RGE} \label{sec:4comp}

The Lagrangian in (\ref{eq:2complag}) can be recast into 4-component form by combining spinors into 4-component Majorana fermions given by $\Psi_{M i}={\left(\psi_M,-\zeta\psi^*_M\right)}_i$ and Dirac fermions, $\Psi_{D i}={\left(\psi_L,-\zeta\psi^*_R\right)}_i$:
\begin{eqnarray}
\nonumber\mathcal{L}_c\ni&&-\left(U^{1a}_{jk}\overline{\Psi}_{Dj}P_L\Psi_{Dk}\Phi_a+U^{2a}_{jk}\overline{\Psi}_{Dj}P_L\Psi_{Dk}\Phi^\dagger_a\right.\\
\nonumber&&+V^a_{jk}\overline{\Psi}_{Dj}P_L\Psi_{Mk}\Phi_a+W^a_{jk}\overline{\Psi}_{Mj}P_L\Psi_{Dk}\Phi^\dagger_a\\
\nonumber&&+\frac{1}{2}X^{1a}_{jk}\overline{\Psi}_{Mj}P_L\Psi_{Mk}\Phi_a+\frac{1}{2}X^{2a}_{jk}\overline{\Psi}_{Mj}P_L\Psi_{Mk}\Phi^\dagger_a\\
&&\left.+\mathrm{h.c.}\right)
\end{eqnarray}
where the scalar $\phi_{a}$ is now complex.

Using this new Lagrangian, the RGEs for $U^{1a}$, $U^{2a}$, $V^{a}$, $W^{a}$, $X^{1a}$ and $X^{2a}$ can be obtained from (\ref{eq:2comprge}). In the case of the MSSM, the Lagrangian does not contain any terms where the Higgs fields ($h$) connect Dirac and Majorana fermions, so $W^{h}$ and $V^{h}$ vanish. Here $h$ can stand for any of the fields $\mathsf{h}$, $\mathcal{H}$, $G^{+}$ or $H^{+}$ since the general form of the RGE is the same for all Higgs fields.

The RGE for $U^{1h}$, which we show as an example, is
\begin{eqnarray}
\nonumber\lefteqn{\left(4\pi\right)^2\left.\frac{d\bu^{1h}}{dt}\right|_{1-loop}=} \qquad \\
\nonumber&&\frac{1}{2}\left[\left(\bu^{1b}\bu^{1b\dagger}+\bu^{2b}\bu^{2b\dagger}+\bv^{b}\bv^{b\dagger}\right)\bu^{1h}\right.\\
\nonumber&&\left.\qquad+\bu^{1h}\left(\bu^{1b\dagger}\bu^{1b}+\bu^{2b\dagger}\bu^{2b}+\bw^{b\dagger}\bw^{b}\right)\right]\\
\nonumber&&+2\left[\bu^{1b}\bu^{2h\dagger}\bu^{2b}+\bu^{2b}\bu^{2h\dagger}\bu^{1b}+\bv^{b}\bx^{2h\dagger}\bw^{b}\right]\\
\nonumber&&+\bu^{1b}\mathrm{Tr}\left\{\left(\bu^{1b\dagger}\bu^{1h}+\bu^{2h\dagger}\bu^{2b}\right)\right.\\
\nonumber&&\left.\qquad\quad\ +\frac{1}{2}\left(\bx^{1b\dagger}\bx^{1h}+\bx^{2h\dagger}\bx^{2b}\right)\right\}\\
\nonumber&&+\bu^{2b}\mathrm{Tr}\left\{\left(\bu^{2b\dagger}\bu^{1h}+\bu^{2h\dagger}\bu^{1b}\right)\right.\\
\nonumber&&\left.\qquad\quad\ +\frac{1}{2}\left(\bx^{2b\dagger}\bx^{1h}+\bx^{2h\dagger}\bx^{1b}\right)\right\}\\
&&-3g^2\left[\bu^{1h}\mathbf{C}^L_2(F)+\mathbf{C}^R_2(F)\bu^{1h}\right],
\end{eqnarray}
where the final term has been constructed by separating $\mathbf{C}_2(F)$ into separate terms for left-handed and right-handed fields. The other RGEs will appear in \cite{abfut}.


\begin{thebibliography}{999}

\bibitem{cas2}
D.J. Casta\~{n}o, E.J. Piard, P. Ramond, Phys. Rev. D \textbf{49}, (1994) 4882

\bibitem{dedes}
A. Dedes, A.B. Lahanas, K. Tamvakis, Phys. Rev. D \textbf{53}, (1996) 3793

\bibitem{martv}
S. Martin, M.T. Vaughn, Phys. Rev. D \textbf{50}, (1994) 2282

\bibitem{cas1}
H. Arason, \textit{et al.}, Phys. Rev. D \textbf{46}, (1992) 3945

\bibitem{HK}
K. Hikasa, M. Kobayashi, Phys. Rev. D \textbf{36}, (1987) 724

\bibitem{machv}
M.E. Machacek, M.T. Vaughn, Nucl. Phys. \textbf{B222}, (1983) 83; Nucl. Phys. \textbf{B236}, (1984) 221; Nucl. Phys. \textbf{B249}, (1985) 70

\bibitem{luo}
M. Luo, H. Wang, Y. Xiao, Phys. Rev. D \textbf(67), (2003) 065019

\bibitem{wss}
H. Baer, X. Tata, \textit{Weak Scale Supersymmetry} (Cambridge, 2006)

\bibitem{abfut}
A. Box, X. Tata (Paper in preparation)

\bibitem{ijet}
F.E. Paige, \textit{et al.}, arXiv:hep-ph/0312045

\bibitem{porod}
W. Porod, T. W\"{o}hrmann, Phys. Rev. D \textbf{55}, (1997) 2907

\end{thebibliography}
\end{document}